\documentclass[aps,prd,notitlepage,superscriptaddress,longbibliography,nofootinbib]{revtex4-1}

\usepackage{dcolumn}    
\usepackage{bm}         
\usepackage{hyperref}   
\usepackage{bmpsize}
\usepackage{url}
\hypersetup{breaklinks=true}
\usepackage[english]{babel}
\usepackage[skip=1pt, justification=raggedright]{caption}
\usepackage{graphicx}
\vspace{-10pt}
\usepackage{amsmath}
\hypersetup{
    colorlinks=true,
    citecolor=blue,
    linkcolor=blue,
    urlcolor=blue
}

\begin{document}

\title{Improving calibration accuracy with torque coupled gravity field calibrator for sub-Hz gravitational wave observation in CHRONOS}

\author{Yuki Inoue}
\thanks{Corresponding author: iyuki@ncu.edu.tw}
\affiliation{Department of Physics,National Central University, Taoyuan, Taiwan}
\affiliation{Center for High Energy and High Field (CHiP), National Central University,  Taoyuan, Taiwan}
\affiliation{Institute of Physics, Academia Sinica, Taipei, Taiwan}
\affiliation{Institute of Particle and Nuclear Studies, High Energy Acceleration Research Organization (KEK), Tsukuba, Japan}

\author{Daiki Tanabe}
\affiliation{Center for High Energy and High Field (CHiP), National Central University,  Taoyuan, Taiwan}
\affiliation{Institute of Physics, Academia Sinica, Taipei, Taiwan}
\affiliation{Institute of Particle and Nuclear Studies, High Energy Acceleration Research Organization (KEK), Tsukuba, Japan}

\author{Vivek Kumar}
\affiliation{Department of Physics,National Central University, Taoyuan, Taiwan}
\affiliation{Center for High Energy and High Field (CHiP), National Central University,  Taoyuan, Taiwan}





\date{\today}

\begin{abstract}

A fundamental challenge in low-frequency gravitational-wave detectors is the limited signal-to-noise ratio (SNR) of calibration lines, particularly in torsion-bar systems where the response is governed by rotational dynamics. In this work, we resolve this issue by optimizing the geometrical configuration of a torque-coupled gravity field calibrator (GCal), achieving an improvement in calibration-line SNR by more than an order of magnitude compared to conventional layouts.

For the Cryogenic sub-Hz cROss torsion-bar detector with quantum NOn-demolition Speed-meter (CHRONOS), the calibration signal appears as a monochromatic line within the $0.1$--$10~\mathrm{Hz}$ band. At $1~\mathrm{Hz}$, the strain-equivalent calibration amplitude reaches $|h_{\rm GCal}| = 1.16 \times 10^{-14}$, corresponding to an SNR density of $|h_{\rm GCal}|/S_h = 4.16 \times 10^{3}$. This demonstrates for the first time that a high-SNR calibration line can be directly injected into the sub-Hz band of a torsion-bar detector.

A first-order perturbative error propagation analysis yields a total fractional systematic uncertainty of $\delta h_{\rm GCal}/h_{\rm GCal} = 0.24\%$, dominated by geometric alignment uncertainties, while contributions from mass uncertainties and the gravitational constant remain subdominant. The corresponding absolute systematic uncertainty is $\delta h_{\rm GCal} \sim 10^{-17}$ at $1~\mathrm{Hz}$.

These results establish torque-coupled gravitational calibration as a practical solution to the longstanding low-SNR problem in sub-Hz torsion-bar detectors and provide a robust pathway toward precision absolute calibration in the low-frequency regime.

\end{abstract}

\maketitle
\section{Introduction}

The first direct detection of gravitational waves (GWs) from a binary
black-hole merger by the Advanced LIGO detectors in 2015~\cite{LIGO2016}
established GW astronomy as a new observational field.
Subsequent observing runs conducted by the global detector network consisting of
LIGO~\cite{Aasi2015_AdvLIGO}, Virgo~\cite{Acernese2015_AdvVirgo}, and KAGRA~\cite{Aso2013_KAGRA_Design} (LVK)
have led to the detection of more than two hundred gravitational-wave events
from compact binary coalescences spanning a wide range of source masses and
distances~\cite{GWTC1,GWTC2,GWTC3,GWTC4}.These observations have enabled precision
tests of general relativity in the strong-field regime, revealed the population
properties of black holes and neutron stars, and opened new avenues for
multi-messenger astronomy and precision cosmology~\cite{Inoue2018_GCal,Campeti2021Measuring}.

Gravitational-wave observations provide a direct measurement of the luminosity
distance to compact binary sources without relying on traditional distance
ladders, making them powerful probes of cosmological parameters such as the
Hubble constant~\cite{Feeney2018_H0,Riess2016_H0,Nissanke2010_GRB, Abbott2017_Nature_H0,Campeti2021Measuring}. As detector sensitivities improve and
event rates increase, systematic uncertainties in detector calibration are
becoming a limiting factor in the accuracy of astrophysical and cosmological
inference. In particular, amplitude calibration uncertainty directly propagates
into the uncertainty of the luminosity distance and therefore into cosmological
measurements. Achieving percent-level accuracy in source parameters, and
ultimately sub-percent constraints on cosmological quantities, requires
calibration precision comparable to the statistical accuracy of future
observations~\cite{Inoue2018_GCal,Campeti2021Measuring, Kuck2009_EUROMET}.

High-precision calibration is therefore essential for maximizing the
scientific return of GW detectors. The strain signal measured
by an interferometer must be reconstructed from the raw photodetector output
through an accurately determined detector response function over the full
observation band. The calibration framework developed for Advanced LIGO has
demonstrated that even small systematic errors can introduce biases in inferred
source parameters and must be carefully tracked and corrected~\cite{Abbott2017Calibration,
Tuyenbayev2016, Cahillane2017_Calibration}. As detector performance approaches design sensitivity and
next-generation detectors are being planned, calibration accuracy has become a
challenge comparable in importance to improving detector sensitivity itself~\cite{Hall2017CalibrationRequirements}.

Several calibration techniques have been developed and deployed in current
detectors. The photon calibrater (PCal)~\cite{Inoue2023_KAGRA_PCal, Goetz2009_LIGO_PCal, Karki2016,Mossavi2006_GEO600_PCal} has served as the
primary calibration method in Advanced LIGO and Virgo, using modulated
radiation pressure to induce a known displacement of the test masses. More
recently, the gravity field calibrater (GCal)~\cite{Inoue2018_GCal, Akutsu2021_KAGRA_Calibration,Matone2007_CQG,Raffai2011_PRD,Oide1980_JJAP} has been proposed and
experimentally demonstrated as an absolute, SI-traceable calibration method
based on the dynamic gravitational field generated by rotating multipole
masses~\cite{Inoue2018_GCal}. Its implementation in the form of the Newtonian
calibrator (NCal)~\cite{Estevez2018NCal,Acernese2018VirgoNCal} has been experimentally demonstrated in Virgo, and further
development and testing are ongoing in KAGRA and LIGO. These developments
indicate that metrology-grade calibration infrastructure is becoming an
essential component of future precision GW detectors~\cite{Aso2013_KAGRA_Design, Akutsu2021KAGRA}.

In contrast, torsion-bar-based GW detectors such as Cryogenic sub-Hz cROss torsion-bar detector with quantum
NOn-demolition Speed-meter (CHRONOS)~\cite{Inoue2025_CHRONOS, Inoue2025_CHRONOS_Optics, Tanabe2025_CHRONOS_Intensity, inoue2026_SPP,inoue2026probingyukawagravitymodulated},
TOBA~\cite{Ando2010}, and TorPeDO~\cite{TorPeDO2019} operate in the sub-Hz frequency band where conventional
calibration approaches face intrinsic limitations. 
The detail of CHRONOS experiment is summarized on CHRONOS science program ~\cite{White_paper_CHRONOS}.
These detectors have
traditionally relied on free-swinging or mechanical excitation methods for
calibration~\cite{Adhikari2003_LIGOCalibration}. However, modern techniques such as photon calibrators or
gravitational field calibrators have not yet been fully implemented in this
class of detectors. One of the primary reasons is that, in conventional
force-coupled configurations, the detector response is significantly
suppressed at low frequencies, resulting in insufficient signal-to-noise ratio
(SNR) for accurate calibration. Nevertheless, these modern calibration methods
offer a crucial advantage: they allow the sensing function to be measured while
the detector remains under closed-loop control, which is indispensable for
achieving high-precision calibration. Establishing a reliable calibration
method in the $0.1$--$10~\mathrm{Hz}$ band for torsion-bar detectors is
therefore a necessary milestone for future low-frequency GW observations.

The CHRONOS aims to extend GW
observations into the $0.1$--$10~\mathrm{Hz}$ frequency range~\cite{Inoue2025_CHRONOS, Inoue2025_CHRONOS_Optics, Tanabe2025_CHRONOS_Intensity}. Achieving its
scientific goals requires unprecedented control of quantum noise,
radiation-pressure noise, and optomechanical couplings in a cryogenic
environment. The speed-meter topology adopted in CHRONOS enables the evasion of
quantum back-action at low frequencies while maintaining stable interferometric
operation~\cite{Wang2013PolarizingSagnac, Chen2003SagnacSRC}. To fully exploit these advantages, calibration methods must be
compatible with the low-frequency sensitivity band, cryogenic operation, and
the triangular Sagnac geometry of the interferometer.

In this paper, we propose a torque-coupled GCal
specifically designed for torsion-bar detectors. The central novelty of this
work is the placement of a rotating quadrupole mass directly beneath the
torsion bar, allowing the gravitational interaction to couple directly to the
rotational degree of freedom of the test mass. This configuration fundamentally
avoids the low-frequency response suppression inherent to conventional
force-coupled GCal schemes and enhances the calibration signal by up to
two orders of magnitude. As a result, practical high-precision calibration at
the $0.1~\mathrm{Hz}$ scale becomes possible for the first time.

In addition, we develop a complete analytical description of the gravitational
torque generated by the rotating quadrupole mass. The multipole expansion is
systematically reformulated by generalizing the expansion coefficients using
binomial coefficients, leading to closed-form expressions that include
higher-order contributions. This analytical framework enables unified and
transparent calculations of detector response and systematic-error propagation,
and provides a general formulation applicable beyond a specific detector
configuration.

These results establish torque-coupled GCal as both an
experimentally feasible and analytically well-defined approach to absolute
calibration in the sub-Hz regime, enabling sub-percent-level amplitude
calibration accuracy required for next-generation torsion-bar GW
detectors such as CHRONOS.

\section{Principle of the GCal}
\label{sec:principle}

\begin{figure}[]
\centering
\includegraphics[width=0.9\linewidth]{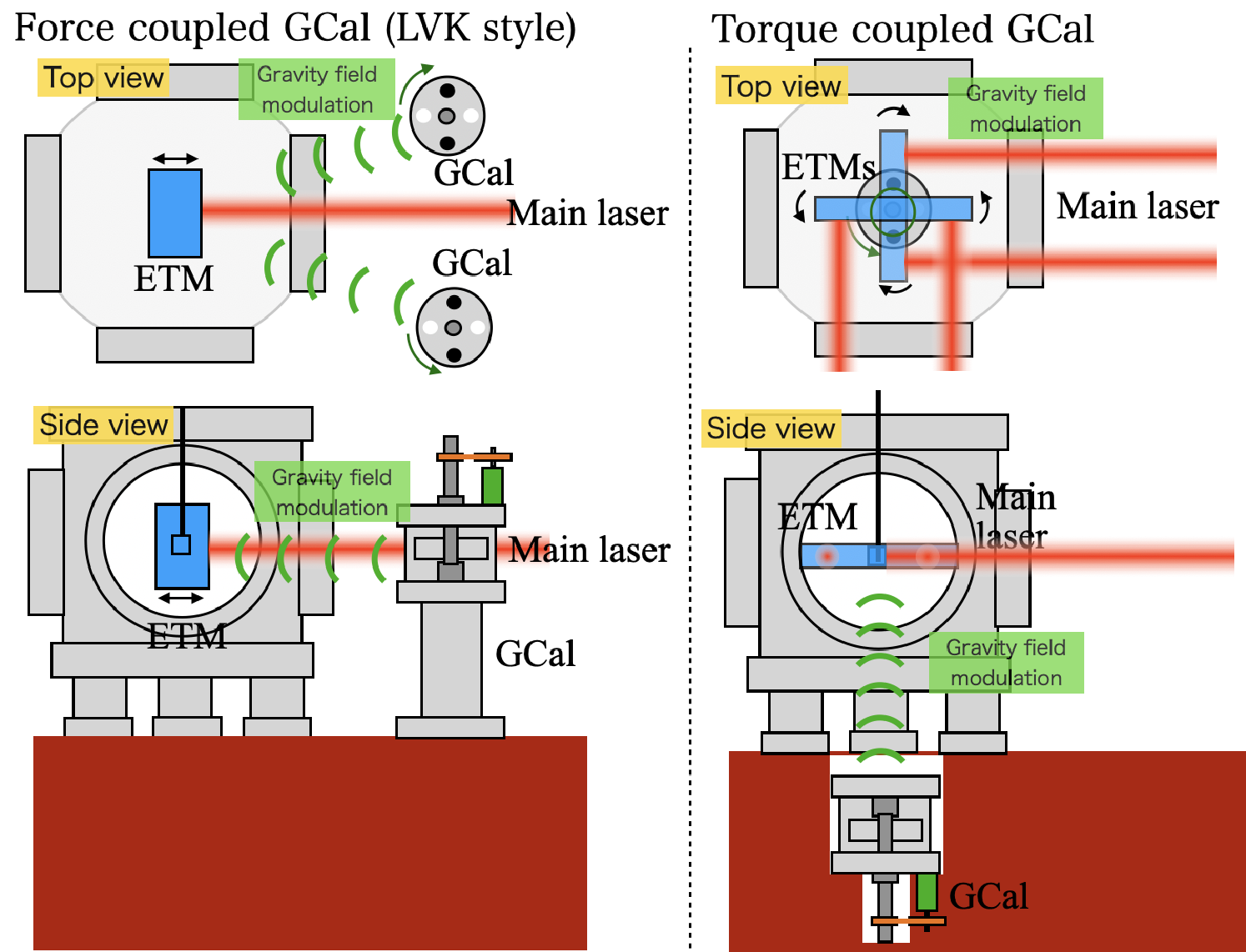}
\caption{
Conceptual comparison between conventional force-coupled GCal and the torque-coupled configuration proposed in this work.
(\textit{Left}) Conventional GCal systems induce calibration through a time-varying gravitational force, producing translational motion of the test mass.
(\textit{Right}) In the torque-coupled configuration for CHRONOS, a rotating quadrupole rotor directly generates a gravitational torque on the torsion-bar test mass, coupling to its rotational degree of freedom.
This direct torque coupling enhances the calibration signal in the sub-Hz band and is the key concept of this work.
}
\label{fig:concept}
\end{figure}

Figure~\ref{fig:concept} shows the conceptual difference between
the conventional force-coupled GCal employed
in kilometer-scale laser interferometers and the torque-coupled
configuration proposed in this work.
In conventional implementations, such as those used in the LVK detectors,
the time-varying gravitational field generated by rotating calibration
masses induces a small translational motion of the test mass~\cite{Inoue2018_GCal, Akutsu2021_KAGRA_Calibration,Estevez2018NCal,Acernese2018VirgoNCal} .
The calibration signal is therefore coupled to the detector output through
displacement.

In contrast, the configuration proposed for CHRONOS places the rotating
quadrupole rotor directly beneath the torsion-bar test mass, allowing the
gravitational interaction to couple directly to the rotational degree of
freedom.
The resulting excitation produces a deterministic gravitational torque
rather than a force, thereby directly driving the fundamental observable
of torsion-bar detectors.
This direct torque coupling constitutes the central idea of this work and
enables a substantial enhancement of the calibration signal in the
sub-Hz frequency band.

\begin{figure}[t]
\centering
\includegraphics[width=0.8\linewidth]{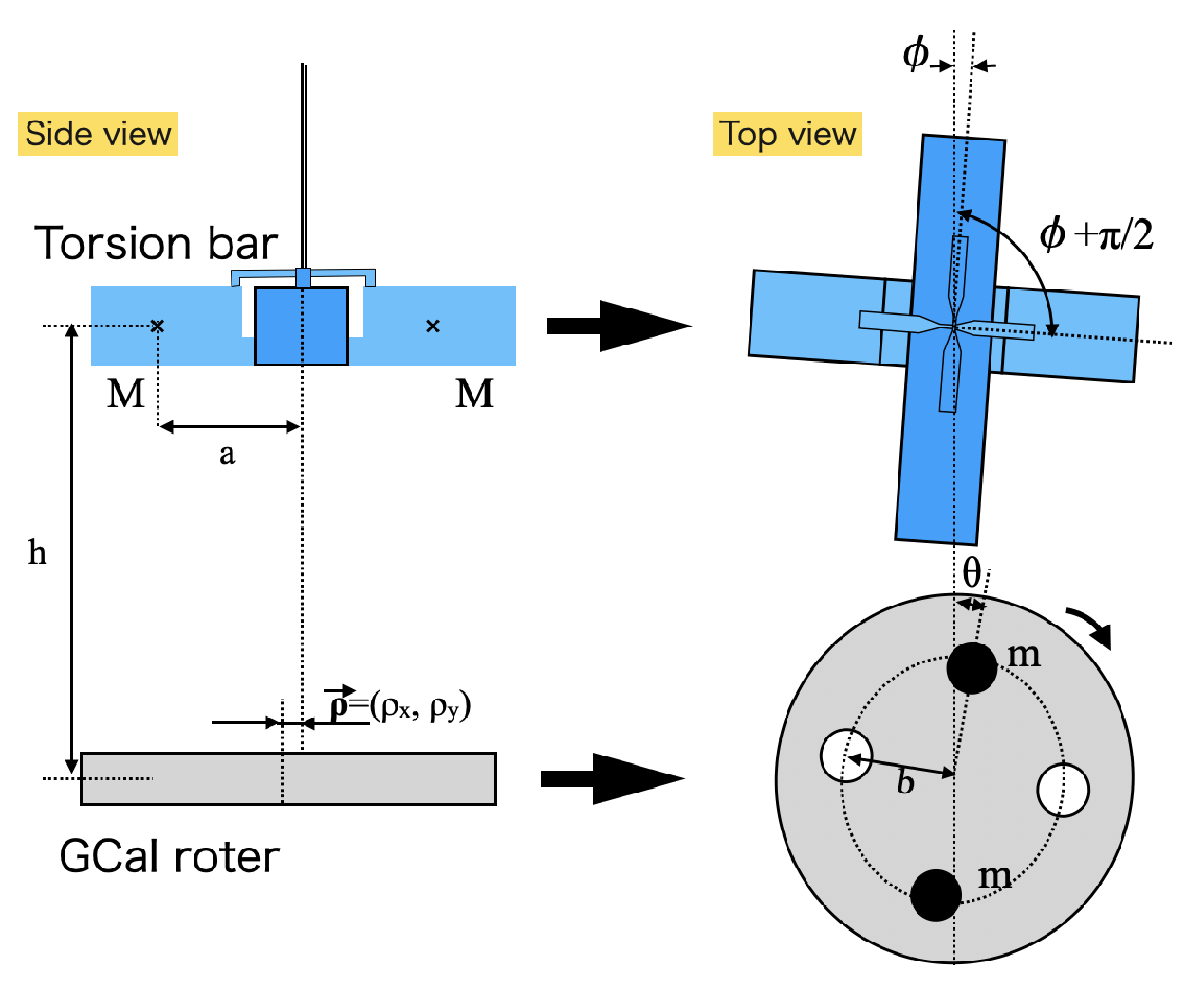}
\caption{
Schematic illustration of the torque-coupled GCal configuration implemented in CHRONOS.
A rotating quadrupole rotor is positioned directly beneath the
torsion-bar test mass.
The time-varying Newtonian gravitational field generated by the rotating
masses produces a periodic torque acting on the torsional degree of
freedom of the bar.
The quadrupole symmetry suppresses lower-order harmonics and generates a
dominant calibration signal at twice the rotation frequency
($2f_{\rm rot}$), enabling a narrow-band and spectrally clean excitation
of the detector response suitable for precision calibration in the
sub-Hz regime.
}
\label{fig:geometory}
\end{figure}

Figure~\ref{fig:geometory} shows the basic configuration of the
GCal adopted for CHRONOS~\cite{Inoue2018_GCal}.
A rotating quadrupole rotor is placed beneath the torsion-bar test mass,
and the time-varying Newtonian gravitational field generated by the rotor
couples directly to the rotational degree of freedom of the suspended bar.
The periodic modulation of the gravitational potential produces a
deterministic torque acting on the torsion bar, resulting in a narrow-band
angular response at twice the rotor rotation frequency.

The concept of the GCal is based on generating a
well-defined gravitational force using moving masses whose geometry and
mass distribution are precisely known~\cite{Inoue2018_GCal,Harms2015}.
Unlike photon calibrators, which rely on radiation pressure and therefore
require accurate knowledge of optical power, beam position, and mechanical
transfer functions, a GCal produces a calibration signal determined by
Newtonian gravity and the geometric configuration of the system.
The resulting calibration line is therefore directly traceable to SI units
through the gravitational constant and the measured mass and distance
parameters. This property makes the GCal particularly attractive as an
absolute calibration reference for precision GW detectors.

The concept of GCal has recently progressed beyond
the proposal stage and is now being actively developed within the
GW detector network.
In particular, the NCal, an implementation of the
GCal concept using rotating masses, has been developed and tested in both
Advanced LIGO and Virgo~\cite{Estevez2018NCal,Acernese2018VirgoNCal}.
The NCal produces a well-characterized time-varying gravitational field
that induces a known displacement of the test masses, providing an
independent cross-check of PCal-based measurements.
These developments demonstrate that gravity-based calibration methods are
becoming an integral component of the calibration strategy for current and
future interferometric detectors.

In the context of torsion-bar detectors, however, the conventional
force-coupled implementation is not optimal.
In force-coupled GCal or NCal schemes, the
time-varying gravitational field induces a small translational motion of
the test mass, which must subsequently couple to the rotational degree of
freedom of the torsion bar.
Since the fundamental observable of torsion-bar detectors is rotational
motion rather than displacement, this indirect coupling is inherently weak.
In systems consisting of two torsion bars, the gravitational coupling to
each bar is not perfectly uniform, and geometric asymmetries introduce
additional corrections that must be accounted for in the calibration model.

Furthermore, the absolute sensitivity of torsion-bar detectors is lower
than that of kilometer-scale laser interferometers, making it difficult to
achieve sufficient SNR for calibration signals in
force-coupled configurations.
In the sub-Hz frequency band, the gravitational response appearing as
translational motion becomes particularly small, and it is therefore
challenging to obtain calibration lines with sufficient strength against
environmental disturbances and control signals.
In practice, calibration methods capable of achieving high SNR at
frequencies around $0.1~\mathrm{Hz}$ have not yet been fully established
even for laser interferometers, and this challenge is more severe for
torsion-bar detectors.
Establishing a calibration scheme that provides sufficient SNR in the
low-frequency regime while remaining compatible with normal interferometer
operation therefore remains an important open problem for future
low-frequency GW observations.

The torque-coupled configuration proposed here is motivated by this
limitation.
By placing the rotating quadrupole rotor directly beneath the torsion bar,
the gravitational interaction couples directly to the torsional mode,
eliminating the need for indirect displacement-to-rotation conversion and
maximizing the calibration signal in the frequency region where the detector
exhibits its highest rotational sensitivity.
This configuration reduces the impact of non-uniform coupling between the
two torsion bars and provides a well-defined deterministic torque acting on
the rotational degree of freedom.

The quadrupole symmetry of the rotor suppresses lower-order harmonics and
generates a dominant signal at $2f_{\rm rot}$, enabling clean spectral
separation from environmental disturbances and control signals.
Because the applied gravitational torque is determined by well-defined
geometric and mass parameters, its amplitude is known a priori.
The observed angular response, on the other hand, probes the mechanical
response of the torsion bar and the overall detector transfer function.
This allows the sensing function and detector response to be measured while
the interferometer remains under normal operating conditions, without
interrupting science observations.
Such capability is essential for achieving sub-percent-level calibration
accuracy in the low-frequency regime targeted by CHRONOS.

\subsection{Torque from gravity-gradient coupling of the GCal}

We begin by defining the geometry of the two-body system.
A Cartesian coordinate system $(x,y,z)$ is introduced such that
the center of the torsion bar coincides with the origin,
and the torsion axis is aligned with the $z$ axis.
The GCal rotor is located below the bar along the negative $z$
direction, and its rotation center is allowed to be laterally
offset from the torsion-bar axis.

The midplane of the torsion bar is located at $z=+h/2$,
while that of the rotor lies at $z=-h/2$,
where $h$ denotes the vertical separation between the two midplanes.
The horizontal offset between the torsion-bar axis and the rotor axis
is described by the vector $\boldsymbol{\rho}=(\rho_x,\rho_y,0)$.

The torsion bar and the rotating quadrupole rotor are each modeled,
to leading order, as two point masses located at their right (R)
and left (L) ends.
The half-length of the torsion bar is denoted by $a$,
so that the total bar length is $2a$.
Similarly, $b$ denotes the radial distance of each rotor mass
from the rotor center.

The parameter $a$ denotes the effective arm length
for gravitational torque coupling.
For a uniform torsion bar, this corresponds to the
distance from the rotation axis to the center of mass
of each half of the bar.
This definition allows the distributed mass of the
torsion bar to be represented by an equivalent
quadrupole mass distribution used in the GCal
torque calculation.

In the analytical model, the torsion bar is therefore represented by two
effective point masses located at the right and left half-bar centers of
mass, separated by a distance
\(2a\).

This definition is adopted because the present calibration analysis
focuses on the leading
\(2f_{\rm rot}\)
component of the gravitational signal.
This component is dominated by the quadrupolar coupling between the
rotating source masses and the effective two-mass representation of the
torsion bar.
The parameter
\(a\)
therefore provides a compact way to describe the leading quadrupole
response relevant to the calibration line.

Higher multipole components associated with the detailed mass
distribution of the torsion bar can also contribute to the gravitational
interaction.
However, these terms mainly appear at higher harmonics of the rotor
frequency and are subdominant for the
\(2f_{\rm rot}\)
calibration signal considered here.

The position vectors of the individual point masses are denoted by
$\mathbf{r}_{\mathrm{bar}}^{R,L}$ and
$\mathbf{r}_{\mathrm{Gcal}}^{R,L}$,
which represent the locations of the right (R) and left (L) masses
of the torsion bar and the GCal rotor, respectively,
measured from the origin of the coordinate system.
Explicitly,
\begin{align}
\mathbf{r}_{\mathrm{bar}}^{R}
  &= \left(a\cos\phi,\; a\sin\phi,\; +\frac{h}{2}\right), \\
\mathbf{r}_{\mathrm{bar}}^{L}
  &= \left(-a\cos\phi,\; -a\sin\phi,\; +\frac{h}{2}\right), \\
\mathbf{r}_{\mathrm{Gcal}}^{R}
  &= \left(\rho_x + b\cos\theta,\;
            \rho_y + b\sin\theta,\;
            -\frac{h}{2}\right), \\
\mathbf{r}_{\mathrm{Gcal}}^{L}
  &= \left(\rho_x - b\cos\theta,\;
            \rho_y - b\sin\theta,\;
            -\frac{h}{2}\right),
\end{align}
where $\phi$ is the azimuthal angle of the torsion bar
and $\theta$ is the rotation angle of the GCal rotor.
Both angles are measured counterclockwise from the $x$ axis
in the $(x,y)$ plane.
The relative angle between the bar and the rotor is defined as
$\Delta = \phi - \theta$.

The gravitational interaction between the moving masses produces
a time-dependent gravity field gradient, which generates a torque
on the torsional degree of freedom of the bar
(see, e.g.,~\cite{Harms2015,Estevez2018NCal}).

For the right--right pair, the separation vector is
\begin{align}
\Delta\mathbf{r}_{RR}
&= \mathbf{r}_{\mathrm{bar}}^{R}
 - \mathbf{r}_{\mathrm{Gcal}}^{R} \nonumber\\
&= \bigl(\mathbf{r}_{\perp}^{R}
          - \boldsymbol{\rho}
          - \mathbf{b}(\theta),\; h\bigr),
\end{align}
where the transverse vectors in the $(x,y)$ plane are defined as
$\mathbf{r}_{\perp}^{R}=a(\cos\phi,\sin\phi,0)$,
and $\mathbf{b}(\theta)=b(\cos\theta,\sin\theta,0)$.
The squared distance is then
\begin{align}
d_{RR}^2
&\equiv \bigl|\Delta\mathbf{r}_{RR}\bigr|^2 \nonumber\\
&= h^2
  + \left|\mathbf{r}_{\perp}^{R}
          - \boldsymbol{\rho}
          - \mathbf{b}(\theta)\right|^2 \nonumber\\
&= h^2+a^2+b^2+\rho_\perp^2
   -2a\,\hat{\mathbf{e}}(\phi)\!\cdot\!\boldsymbol{\rho}
   +2b\,\hat{\mathbf{e}}(\theta)\!\cdot\!\boldsymbol{\rho}
   -2ab\cos\Delta ,
\end{align}
where $\rho_\perp^2\equiv \rho_x^2+\rho_y^2$ and
$\hat{\mathbf{e}}(\alpha)\equiv(\cos\alpha,\sin\alpha,0)$.

It is convenient to separate the angle-independent geometric terms
from the angular dependence.
We define
\begin{equation}
R_0^2 \equiv a^2 + b^2 + h^2 + \rho_\perp^2 ,
\label{eq:R_zero}
\end{equation}
where $\rho_\perp^2 \equiv \rho_x^2+\rho_y^2$.
The projections of the offset vector onto the bar and rotor directions
are written as
$C_\phi \equiv \hat{\mathbf e}(\phi)\!\cdot\!\boldsymbol{\rho}$
and
$C_\theta \equiv \hat{\mathbf e}(\theta)\!\cdot\!\boldsymbol{\rho}$.

The four endpoint separations can then be written compactly as
\begin{align}
d_{RR}^2 &= R_0^2 -2a C_\phi +2b C_\theta -2ab\cos\Delta, \\
d_{RL}^2 &= R_0^2 -2a C_\phi -2b C_\theta +2ab\cos\Delta, \\
d_{LR}^2 &= R_0^2 +2a C_\phi +2b C_\theta +2ab\cos\Delta, \\
d_{LL}^2 &= R_0^2 +2a C_\phi -2b C_\theta -2ab\cos\Delta .
\end{align}

The total Newtonian potential is obtained by summing over the four
point-mass pairs formed between the torsion bar and the GCal rotor.
Each object is modeled as two identical endpoint masses, where $M$
denotes the mass located at each end of the torsion bar and $m$ the
mass at each end of the GCal rotor. Here, $M$ corresponds to one half
of the total mass of the torsion bar. In order to account for the
quadrupole mass distribution, we approximate the torsion bar by
assuming that the mass is effectively localized at the centers of mass
of its right and left halves. Since the present calculation focuses on
the $2f_{rot}$ component of the signal, this quadrupole approximation provides
a valid and sufficiently accurate description of the system.

The total potential is therefore written as
\begin{align}
U(\phi,\theta)
&= -G mM
\left(
\frac{1}{d_{RR}}+\frac{1}{d_{RL}}
+\frac{1}{d_{LR}}+\frac{1}{d_{LL}}
\right),
\end{align}
where gravity constant is defined as $G$. 
In the following, we assume a hierarchy of length scales
$\rho \ll a,\; b \ll R_0$,
where $2a$ and $2b$ are the effective lengths of the torsion bar and the radius of GCal rotor masses,
respectively, and $\boldsymbol{\rho}$ represents a small lateral offset.
The interaction can then be expanded perturbatively around the
reference separation $R_0$, following the standard treatment of
gravity-gradient interactions generated by moving masses
(see, e.g.,~\cite{Harms2015}).

To organize the perturbative expansion in the presence of a lateral
offset, we define the dimensionless quantities
\begin{equation}
\alpha \equiv \frac{2a}{R_0^2} C_\phi,
\qquad
\beta  \equiv \frac{2b}{R_0^2} C_\theta,
\qquad
\xi \equiv \frac{2ab}{R_0^2}.
\end{equation} \label{eq:approx}

Under the assumed hierarchy,
\[
\alpha=\mathcal{O}\!\left(\frac{a\rho}{R_0^2}\right),
\qquad
\beta=\mathcal{O}\!\left(\frac{b\rho}{R_0^2}\right),
\qquad
\xi=\mathcal{O}\!\left(\frac{ab}{R_0^2}\right),
\]
we have
\(
\alpha,\beta \ll \xi
\)
for
\(
\rho \ll a,b
\).
The quantities $\alpha$ and $\beta$ therefore represent
offset-suppressed corrections to the leading interaction.

The perturbative ordering scheme, cancellation of the linear
offset-dependent terms, and numerical validation of the neglected
corrections are summarized in
Appendix~A.

With these definitions, the inverse separations can be written as
\begin{equation}
\frac{1}{d_{ij}}
=
\frac{1}{R_0}(1+x_{ij})^{-1/2},
\end{equation}
where
\begin{align}
x_{RR}&=-\alpha+\beta-\xi\cos\Delta,
&
x_{RL}&=-\alpha-\beta+\xi\cos\Delta,
\nonumber\\
x_{LR}&=+\alpha+\beta+\xi\cos\Delta,
&
x_{LL}&=+\alpha-\beta-\xi\cos\Delta .
\end{align}

Since $|x_{ij}|\ll1$, the inverse distance can be expanded using
the binomial series
\begin{equation}
(1+x)^{-1/2}
=
\sum_{k=0}^{\infty}
\binom{2k}{\,k}
\frac{(-1)^k}{4^k}
x^k,
\label{eq:binomial_series}
\end{equation}
allowing the Newtonian potential to be expressed systematically
as a perturbative series in the small parameters
$\xi$, $\alpha$, and $\beta$.

Under the hierarchy
\(
\rho \ll a,b \ll R_0
\),
the offset-dependent terms are parametrically suppressed.
Neglecting the offset-suppressed contributions beyond linear order in
$\alpha$ and $\beta$ (which cancel in the symmetric sum),
we set
\(
\alpha=\beta=0
\)
in the leading-order potential.
Then
\begin{align}
U(\phi,\theta)
&\simeq
-\frac{GmM}{R_0}
\left[
2(1-\xi\cos\Delta)^{-1/2}
+
2(1+\xi\cos\Delta)^{-1/2}
\right]
\nonumber\\
&=
-\frac{2GmM}{R_0}
\left[
(1-\xi\cos\Delta)^{-1/2}
+
(1+\xi\cos\Delta)^{-1/2}
\right].
\end{align}

Using Eq.~(\ref{eq:binomial_series}), we obtain
\begin{align}
&(1+\xi\cos\Delta)^{-1/2}
+
(1-\xi\cos\Delta)^{-1/2}
\nonumber\\
&=
\sum_{k=0}^{\infty}
\binom{2k}{\,k}
\frac{(-1)^k}{4^k}
\left[
(\xi\cos\Delta)^k
+
(-\xi\cos\Delta)^k
\right]
\nonumber\\
&=
2
\sum_{n=0}^{\infty}
\binom{4n}{\,2n}
\frac{1}{16^n}
\xi^{2n}\cos^{2n}\Delta ,
\label{eq:even_expansion}
\end{align}
where the odd-order terms cancel identically, and the remaining
even-order contributions are obtained by setting
\(
k=2n
\).

Therefore, the Newtonian potential becomes
\begin{equation}
U(\phi,\theta)
\simeq
-\frac{4GmM}{R_0}
\sum_{n=0}^{\infty}
\binom{4n}{\,2n}
\frac{1}{16^n}
\xi^{2n}
\cos^{2n}\Delta .
\end{equation}

Equivalently,
\begin{equation}
U(\phi,\theta)
\simeq
-\frac{4GmM}{R_0}
\sum_{n=0}^{\infty}
\binom{4n}{\,2n}
\frac{1}{4^n}
\left(
\frac{ab}{R_0^2}
\right)^{2n}
\cos^{2n}\Delta .
\label{eq:newtonian_cos_series}
\end{equation}

The angular dependence therefore appears only through even powers of
$\cos\Delta$, reflecting the left--right symmetry of the mass
configuration.

This formulation separates the calibration signal at
$2f_{\rm rot}$
from the remaining contributions without invoking an additional
small-angle approximation, and remains valid in the presence of a finite
lateral offset.
In the symmetric limit of vanishing offset, only even harmonics survive,
and the dominant oscillatory contribution arises from the term
proportional to
$\cos(2\Delta)$,
which provides the calibration line used in the experiment.

To extract the harmonic structure explicitly, we use the Fourier
decomposition
\begin{equation}
\cos^{2n}\Delta
=
\frac{1}{2^{2n}}
\left[
\binom{2n}{\,n}
+
2\sum_{j=1}^{n}
\binom{2n}{\,n-j}
\cos(2j\Delta)
\right].
\label{eq:cos_fourier}
\end{equation}

The Fourier components therefore appear only at even harmonics of
$\Delta$, reflecting the symmetry
\(
\Delta \rightarrow -\Delta
\).

In particular, the coefficient of the
$\cos(2\Delta)$
component is obtained from the
$j=1$
term,
\begin{equation}
\cos^{2n}\Delta
\supset
\frac{1}{2^{2n-1}}
\binom{2n}{\,n-1}
\cos(2\Delta).
\label{eq:cos2_component}
\end{equation}

we see that the gravitational interaction produces only even harmonics.
This reflects the left--right symmetry of the mass configuration and is
a general property of gravity-gradient interactions generated by
quadrupole mass distributions.
In particular, the term proportional to $\cos(2\Delta)$ gives the $2f$
response that is used as the calibration signal in GCal and NCal
implementations~\cite{Inoue2018_GCal,Estevez2018NCal}.

Differentiating the potential with respect to $\phi$ yields the torque
acting on the torsion bar,
\begin{equation}
\tau_\phi
=
-\frac{\partial U}{\partial \phi}.
\end{equation}

Since the potential is written as a series in $\cos^{2n}\Delta$,
we extract the $2f_{\rm rot}$ component using
Eq.~(\ref{eq:cos2_component}).  The corresponding part of the potential is
\begin{equation}
U_{2f}(\phi,\theta)
\simeq
-\frac{4GmM}{R_0}
\sum_{n=1}^{\infty}
\binom{4n}{\,2n}
\binom{2n}{\,n-1}
\frac{1}{2^{6n-1}}
\xi^{2n}
\cos(2\Delta).
\end{equation}

Using
\begin{equation}
\frac{\partial}{\partial\phi}\cos(2\Delta)
=
-2\sin(2\Delta),
\qquad
\Delta=\phi-\theta,
\end{equation}
we obtain
\begin{equation}
\tau_\phi^{(2f)}
=
-\frac{GmM}{R_0}
\sum_{n=1}^{\infty}
\binom{4n}{\,2n}
\binom{2n}{\,n-1}
\frac{1}{2^{6n-4}}
\xi^{2n}
\sin(2\Delta).
\label{eq:analytical_formula}
\end{equation}

The GCal rotor consists of two identical mass clumps placed
diametrically opposite to each other.
This two-clump symmetry eliminates all odd harmonics of the rotation
frequency, since contributions proportional to
$\sin[(2k+1)\Delta]$ cancel between the two clumps.
Consequently, only even harmonics remain, and the torque is purely
proportional to $\sin(2\Delta)$, consistent with the quadrupole nature
of the gravitational excitation~\cite{Harms2015}.

The torque can be written in a compact form as
\begin{equation}
\tau_\phi^{(2f)}(t)
=
-\tau_2 \sin(2\Delta),
\end{equation}
where
\begin{equation}
\tau_2
=
\frac{GmM}{R_0}
\left[
\frac{3}{2}\,\xi^2
+\frac{35}{32}\,\xi^4
+\frac{3465}{4096}\,\xi^6
+\cdots
\right]. \label{eq:tau2_expand}
\end{equation} 

The CHRONOS detector employs two orthogonal torsion bars aligned
along the $X$ and $Y$ axes.
Their orientations are defined by $\phi_X=0$ and
$\phi_Y=\frac{\pi}{2}$.
For a rotor angle $\theta=\omega t$, the phase differences become
$\Delta_X=-\omega t$ and
$\Delta_Y=\frac{\pi}{2}-\omega t$, respectively.
The torques acting on the two torsion bars are therefore

\begin{align}
\tau_X(t)
&=
\eta_{GCAL}\,\tau_2\,\sin(2\omega t),\nonumber\\
\tau_Y(t)
&=
-\eta_{GCAL}\,\tau_2\,\sin(2\omega t).
\end{align}

Thus, the two orthogonal torsion bars experience equal-amplitude torques
with opposite signs. This quadrature symmetry allows the differential
readout to isolate the calibration signal at $2f_{\rm rot}$ while
suppressing common-mode contributions.

The factor \(\eta_{\rm GCal}\) represents the finite-width correction
associated with the effective transverse extent of the torsion-bar mass
distribution. In the ideal thin-bar limit, one has
\(\eta_{\rm GCal}=1\), whereas the finite transverse width reduces the
effective quadrupole response.

Using the multipole expansion up to \(\mathcal O(\xi^6)\), we obtain
\begin{equation}
\eta_{\rm GCal}=0.974
\end{equation}
for the present detector geometry.
The detailed derivation and higher-order evaluation are presented in
Appendix~B.

The correction factor \(\eta_{\rm GCal}\) accounts for the dominant
finite-size and finite-thickness effects of the torsion-bar mass
distribution. Consequently, the uncertainty assigned to the effective
arm length \(a\) in Table~\ref{tab:gcal_params} represents only the geometrical uncertainty
associated with determining the equivalent quadrupole arm length after
applying this correction. No additional independent uncertainty is
assigned to \(\eta_{\rm GCal}\), since the multipole expansion and
numerical evaluation presented in Appendix~B are considered sufficiently
accurate for the present analysis.

\subsection{Equation of Motion}

To relate the external torque to the angular response of each torsion bar,
we model each bar as a single-degree-of-freedom torsional oscillator with
moment of inertia $I$, torsional spring constant $\kappa$, and damping
coefficient $\Gamma$.
The equation of motion under an external torque $\tau_\phi(t)$ is
\begin{equation}
I\,\ddot{\phi}(t)+\Gamma\,\dot{\phi}(t)+\kappa\,\phi(t)
=\tau_\phi(t),
\label{eq:eom_basic}
\end{equation}
which corresponds to the standard equation for a damped torsional
oscillator used in precision mechanical measurements and
GW detectors~\cite{Saulson1990}.

The resonance frequency $\omega_0$ and quality factor $Q$ are defined by
$\omega_0^2=\kappa/I$ and $\Gamma=I\omega_0/Q$.
Eq.~\eqref{eq:eom_basic} then becomes
\begin{equation}
\ddot{\phi}(t)+\frac{\omega_0}{Q}\dot{\phi}(t)
+\omega_0^2\phi(t)
=\frac{\tau_\phi(t)}{I}.
\label{eq:eom_reduced}
\end{equation}

The response frequency of the interferometer, $\Omega$, is related to the
rotor angular frequency $\omega$ through
$\Omega = 2\omega$, since the quadrupole gravitational potential generated by the rotating
calibrator produces a torque at twice the rotor rotation frequency.

Moving to the frequency domain, the equation of motion becomes
\begin{equation}
\left(-I\Omega^2+i\Gamma\Omega+\kappa\right)\Phi(\Omega)
=\tau(\Omega),
\end{equation}
yielding the angle--torque transfer function (mechanical susceptibility)
\begin{equation}
\chi_{exact}(\Omega)
\equiv
\frac{\Phi(\Omega)}{\tau(\Omega)}
=
\frac{1}{I\left(\omega_0^2-\Omega^2
+i\,\frac{\omega_0\Omega}{Q}\right)}.
\label{eq:chi_general}
\end{equation}

Since the GCal torque consists of a single harmonic at $2\omega$, the
steady-state angular responses of the two torsion bars are
$\Phi_X(\Omega)=\chi_{exact}(\Omega)\eta_{GCAL} \tau_2$ and
$\Phi_Y(\Omega)=-\chi_{exact}(\Omega)\eta_{GCAL}\tau_2$.

The observable quantity in CHRONOS is the differential angle between the
two orthogonal torsion bars,
$\phi_{\rm diff}(t)\equiv[\phi_X(t)-\phi_Y(t)]$,
for which the frequency-dominant response amplitude becomes
\begin{equation}
\Phi_{\rm diff}(\Omega)
=2\chi_{exact}(\Omega)\eta_{GCAL}\tau_2.
\label{eq:phi_diff}
\end{equation}

In order to relate the measured angular response to an equivalent
gravitational-wave strain, we introduce the antenna response of the
torsion-bar configuration.
In the long-wavelength limit, the differential angular response of two
orthogonal bars can be written as
\begin{equation}
\Phi_{\rm diff}(\Omega)
= \eta_g \left( h_{+}F_{+} + h_{\times}F_{\times} \right),
\label{eq:antenna_response}
\end{equation}
where $\eta_g \equiv I_{\rm eff}/I = 0.936$ is the geometrical coupling factor,
and $F_{+}$ and $F_{\times}$ are the antenna pattern functions.
For normal incidence along the $z$ axis, these are given by
\begin{equation}
F_{+}=\sin 2\alpha, \qquad
F_{\times}=-\cos 2\alpha,
\end{equation}
with $\alpha$ denoting the angle between the bar axis and the principal
polarization axis of the gravitational wave.

Using this relation, the angular response induced by the gravitational
calibrator can be expressed in terms of a strain-equivalent amplitude.
We therefore define the strain-equivalent calibration signal as
\begin{equation}
h_{\rm GCal}(\Omega)
\equiv
\frac{\Phi_{\rm diff}(\Omega)}
{\eta_g\,|F_{\rm eff}|},
\label{eq:h_gcal}
\end{equation}
where
$|F_{\rm eff}| \equiv \sqrt{F_{+}^{2}+F_{\times}^{2}}$
denotes the effective antenna response.
This definition allows the calibration-line amplitude produced by the
gravitational calibrator to be directly compared with the detector strain
sensitivity.

For the purpose of estimating representative calibration amplitudes,
we assume normal incidence in the long-wavelength limit.
In this configuration, the antenna factors satisfy
$F_{+}^{2}+F_{\times}^{2}=1$, and therefore
$|F_{\rm eff}|=1$.
The strain-equivalent calibration amplitude is thus evaluated using
the optimal antenna response.

\subsection{High-frequency approximation}

In CHRONOS, the torsion-bar resonance frequency is typically
$\omega_0 < 10^{-2}\,\mathrm{Hz}$, whereas the gravitational calibrator
(GCal) is operated in the range
$2\omega \sim 0.1$--$10\,\mathrm{Hz}$.
The calibration frequency therefore always satisfies
$\Omega = 2\omega \gg \omega_0$.
In this regime, the mechanical response of the torsion-bar system
simplifies considerably.

Because the inertial term dominates the denominator of
Eq.~\eqref{eq:chi_general}, the mechanical susceptibility reduces to
\begin{equation}
\chi_{HF}(\Omega) \simeq -\frac{1}{I\Omega^2}.
\end{equation}  \label{eq:approx_HF}

Using Eq.~\eqref{eq:phi_diff} together with the definition of the
strain-equivalent calibration signal introduced in
Eq.~\eqref{eq:h_gcal}, the GCal-induced strain amplitude becomes
\begin{equation}
h_{\rm GCal}(\Omega)
=
\frac{2\eta_{GCAL}\,\chi_{HF}(\Omega)\tau_2}
{\eta_g |F_{\rm eff}|}.
\end{equation}
Applying the high-frequency approximation yields
\begin{equation}
h_{\rm GCal}(\Omega)
\simeq
-\frac{2\, \eta_{GCAL} \tau_2}
{\eta_g |F_{\rm eff}|\, I \Omega^2}.
\label{eq:h_gcal_highf}
\end{equation}

Thus, in the frequency range relevant for CHRONOS, the calibration signal
exhibits a characteristic $1/\Omega^2$ dependence determined purely by
the inertia of the torsion bars.
The resulting strain-equivalent response is directly related to the
known gravitational torque generated by the calibration masses.
Because the response is governed primarily by inertia rather than by
uncertainties in the torsional spring constant or damping parameters,
the $2f_{\rm rot}$ calibration line provides a clean and robust absolute
reference for calibrating the detector strain response.
The applicability of the high-frequency approximation used in this
analysis is discussed in Appendix~C.

\section{Experimental configuration and sensitivity}
\label{sec:exp_config}

In this section, we evaluate the performance of the proposed
GCal scheme under realistic experimental conditions
of the CHRONOS.
The objective is to demonstrate that the gravitationally induced
calibration signal can be clearly resolved above the intrinsic detector
noise while remaining compatible with continuous detector operation.
This requires a direct comparison between the predicted GCal-induced
response and the differential angular sensitivity of the interferometer.

The CHRONOS employs a torsion-bar interferometric configuration
in which a pair of massive sapphire bars are suspended as a low-frequency rotational
resonator operating less than $\omega_0/2\pi \simeq 0.01\,\mathrm{Hz}$.
The GCal is installed directly beneath the
torsion bar and provides a precisely calculable gravitational torque at
twice the rotor frequency.
The quadrupole rotor consists of two point-like calibration masses
mounted at radius $b = 0.16\,\mathrm{m}$, while the effective arm length
of the torsion bar is $a = 0.302\,\mathrm{m}$.
The vertical separation between the rotor plane and the torsion-bar
midplane is $h = 2.00\,\mathrm{m}$.
These values correspond to the actual CHRONOS geometry and are
used consistently in the torque-series computation and the sensitivity
evaluation presented below~\cite{Inoue2025_CHRONOS_Optics}.

The effective arm length $a$, introduced in Sec.~\ref{sec:principle} as the
effective lever arm for gravitational torque coupling,
can be derived by approximating the mass distribution of
the torsion bar as an equivalent quadrupole mass
distribution with respect to the rotational axis.
In this description, the continuous mass distribution of the torsion bar
is replaced by an equivalent configuration that reproduces the same
quadrupole moment relevant for the rotational degree of freedom.
Since the present analysis focuses exclusively on the calibration signal
at twice the rotor rotation frequency ($2f_{\rm rot}$), it is sufficient
to retain only the dominant quadrupole contribution in the gravitational
interaction.
Under this condition, the quadrupole approximation provides an adequate
description of the coupling responsible for the $2f$ response.
Higher-order multipole contributions introduce only small corrections to
the calibration amplitude and do not significantly affect the evaluation
of the SNR or the calibration accuracy considered in
this work.
The quadrupole approximation therefore provides a consistent and
physically well-motivated framework for the torque-series expansion and
the sensitivity evaluation presented in the following sections.

Table~\ref{tab:gcal_params} summarizes the parameters adopted in the
GCal model.
The listed uncertainties correspond to realistic accuracies
for mass, geometry, and alignment parameters and are later used in the
systematic-error evaluation of the calibration accuracy.

\begin{table}[t]
\centering
\begin{tabular}{c|c|c}
\hline
& \textbf{Values} & \textbf{Relative uncertainty} \\
\hline
$G$
& $6.67430(15) \times 10^{-11}\,\mathrm{m^3\,kg^{-1}\,s^{-2}}$
& $0.00225~\%$ \\

$m$
& $1.88\,\mathrm{kg}$
& $0.00531~\%$ \\

$M$
& $85.5\,\mathrm{kg}$
& $0.000165~\%$ \\

$I$
& $19.9\,\mathrm{kg\,m^2}$
& $0.00296~\%$ \\

$h$
& $2.00\,\mathrm{m}$
& $0.050~\%$ \\

$a$
& $0.302\,\mathrm{m}$
& $0.00331~\%$ \\

$b$
& $0.160\,\mathrm{m}$
& $0.00884~\%$ \\

$\rho_\perp$
& $0.001\,\mathrm{m}$
& $100~\%$ \\
\hline
\end{tabular}
\caption{
Summary of the parameters used in the GCal model.
The rotor mass $m$ listed here corresponds to the tungsten configuration
used as the reference case throughout the analysis.
For alternative rotor materials, including SUS304 stainless steel and
A5083 aluminum alloy, the corresponding mass values and derived
calibration parameters are summarized separately in
Table~\ref{tab:gcal_material_comparison}.
}
\label{tab:gcal_params}
\end{table}

The interpretation of calibration signals in the presence of such noise
sources follows the standard framework developed for precision
interferometric GW detectors
~\cite{Saulson1990,Harms2015,Abbott2017Calibration}.
Because the GCal produces a deterministic gravitational excitation,
the detectability of the calibration signal is determined by its
amplitude relative to the noise spectral density at the excitation
frequency.

\subsection{Sensitivity and GCal response}

\begin{figure}[t]
\centering
\includegraphics[width=0.85\linewidth]
{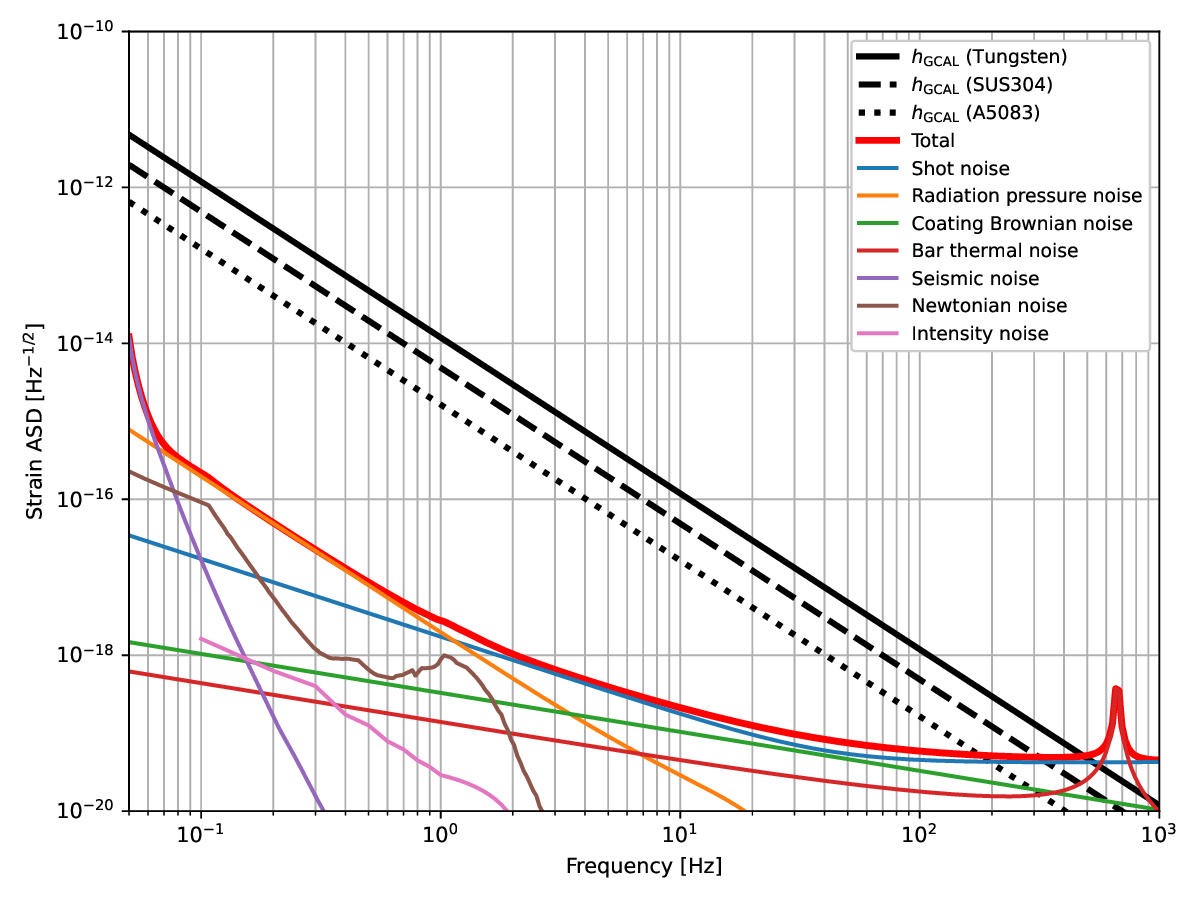}
\caption{
Calculated strain sensitivity of the CHRONOS
(red curve) together with the predicted GCal
response.
The colored curves show the expected GCal-induced differential angular
motion for calibration rotors made of tungsten (black solid),
SUS304 stainless steel (black dashed), and aluminum alloy A5083 (black dotted),
assuming identical rotor geometry.
Because the gravitational torque scales linearly with the calibration
mass, the response amplitude follows the material density.
The comparison illustrates the achievable calibration margin for
different practical material choices without modifying the mechanical
configuration of the detector.
}
\label{fig:sens_curve}
\end{figure}

Figure~\ref{fig:sens_curve} shows the calculated strain
sensitivity of the CHRONOS together with the predicted
GCal-induced response.
The black curve represents the equivalent differential angular noise
spectral density of the detector, including contributions from quantum
noise, suspension thermal noise, seismic coupling, and gravity-gradient
noise~\cite{Inoue2025_CHRONOS, Inoue2025_CHRONOS_Optics, Tanabe2025_CHRONOS_Intensity}.
 The detail of noise model and sensitivity is discussed in Inoue and Tanabe {\it et al.} \cite{Inoue2025_CHRONOS, Inoue2025_CHRONOS_Optics, Tanabe2025_CHRONOS_Intensity}.

The solid, dashed and dotteded lines show the expected differential angular response
induced by the GCal, computed using the higher-order
gravitational torque series combined with the mechanical transfer
function of the torsion-bar resonator.
For identical rotor geometries, the gravitational torque scales linearly
with the calibration mass, and therefore the response amplitude follows
the density of the rotor material.

Because the GCal excitation is monochromatic, the calibration signal
appears as a spectral line at $2f_{\rm rot}$.
In contrast to broadband signals, the detectability of such a line is
determined by the noise spectral density evaluated at the excitation
frequency rather than by an integrated sensitivity over frequency.
This principle is widely employed in PCals and NCals in current interferometric detectors
~\cite{Karki2016,Estevez2018NCal}.
Consequently, even a relatively small excitation torque can yield a high
effective SNR when the excitation frequency lies
within a low-noise region of the detector band.

In the CHRONOS configuration, the calibration band of
$0.1$--$10 \mathrm{Hz}$ lies well above the torsional resonance and below
the frequency range where quantum shot noise dominates.
In this regime the torsion bars behave as inertial elements, producing a
well-defined $1/\Omega^2$ mechanical response.
The calibration signal is therefore directly proportional to the applied
gravitational torque, enabling an absolute conversion between the known
excitation torque and the differential angular response.

\begin{figure}[t]
\centering
\includegraphics[width=0.85\linewidth]
{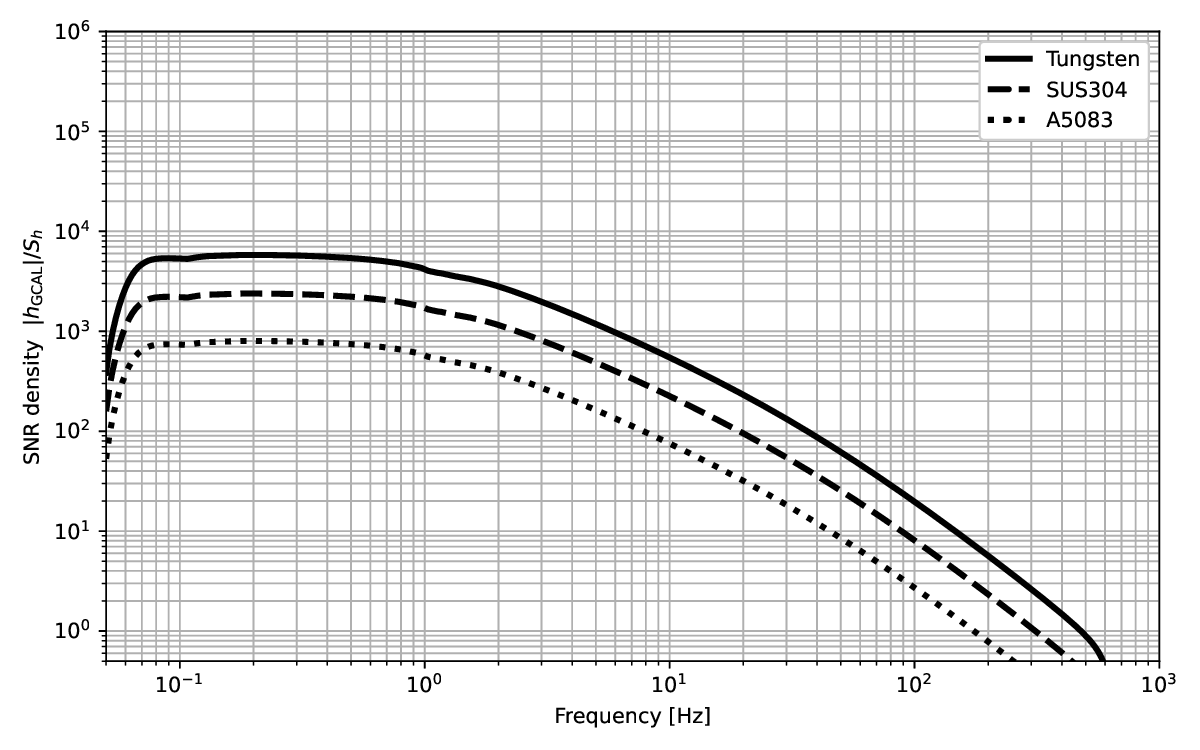}
\caption{
SNR density of the GCal-induced calibration signal
for the CHRONOS.
The curves correspond to calibration rotors made of tungsten (solid),
SUS304 stainless steel (dashed), and aluminum alloy A5083 (dotted).
Because the GCal excitation is monochromatic, the SNR is determined by
the noise spectral density at the excitation frequency.
The large SNR observed in the sub-Hz band originates from the
combination of line excitation and the $1/\Omega^2$ inertial response of
the torsion-bar mode.
}
\label{fig:gcal_snr_density}
\end{figure}

Because the GCal applies equal and opposite torques to the two orthogonal
torsion bars, the calibration signal appears purely in the differential
channel, while common-mode contributions are suppressed by symmetry.
As shown in Fig.~\ref{fig:sens_curve}, the calibration line remains well
above the intrinsic detector noise across the operational band,
demonstrating that the sensing function can be measured with high
statistical precision.
To quantify the calibration performance more directly,
Fig.~\ref{fig:gcal_snr_density} shows the corresponding SNR density.
For monochromatic excitation, the SNR is determined directly by the ratio
between the signal amplitude and the noise spectral density at the
excitation frequency, without requiring frequency integration.
The SNR density therefore provides a direct measure of the statistical
significance of the calibration signal as a function of frequency.
The combined effect of the low-noise region in the detector band and the
$1/\Omega^2$ inertial response of the torsion-bar mode leads to a very
large SNR in the sub-Hz regime.

Using the estimated sensitivity of CHRONOS~\cite{Inoue2025_CHRONOS,
Inoue2025_CHRONOS_Optics, Tanabe2025_CHRONOS_Intensity},
the calibration-line amplitude is evaluated relative to the total noise
amplitude spectral density.
The signal-to-noise ratio density is defined as
\[
\mathrm{SNR}(\Omega) =
\frac{|h_{\rm GCal}(\Omega)|}{S_{\rm tot}(\Omega)},
\]
where $S_{\rm tot}$ denotes the total angular noise ASD of the detector.

At $1\,\mathrm{Hz}$, the predicted SNR density reaches
\[
\mathrm{SNR}(1\,\mathrm{Hz}) = 4.16\times10^{3}
\]
for a tungsten calibration rotor.
For identical geometrical configurations, the corresponding values are
$1.72\times10^{3}$ for SUS304 stainless steel and
$5.73\times10^{2}$ for aluminum alloy A5083.
The absolute calibration-line amplitudes at $1\,\mathrm{Hz}$ are
$1.16\times10^{-14}\,\mathrm{rad}$,
$4.75\times10^{-15}\,\mathrm{rad}$, and
$1.59\times10^{-15}\,\mathrm{rad}$, respectively.

These results demonstrate that the calibration-line amplitude, and hence
the SNR density, scales approximately linearly with the rotor mass and
therefore with the material density for a fixed geometry.
Even for the lowest-density material considered (A5083), the calibration
line remains well above the total noise level around $1\,\mathrm{Hz}$,
providing a substantial calibration margin without requiring any
modification of the mechanical configuration of the detector.
The assumed values and calculation results are summarized in
Table~\ref{tab:gcal_material_comparison}.

\begin{table}[t]
\centering
\caption{
Comparison of three rotor materials assuming identical rotor
geometry and volume.
The table summarizes the rotor endpoint mass,
material density, absolute calibration-line amplitude at
$1\,\mathrm{Hz}$, signal-to-noise ratio density at
$1\,\mathrm{Hz}$, and the resulting systematic uncertainties.
The relative systematic error corresponds to
$\sigma_{\rm sys}/|\Phi_{\rm diff}|$.
}
\label{tab:gcal_material_comparison}

\scriptsize

\begin{tabular}{lcccccc}
\hline
Material
& Mass
& Density
& Amplitude
& SNR
& Sys. err.
& Rel. sys.
\\

& [kg]
& [kg\,m$^{-3}$]
& [rad]
& @1 Hz
& [rad]
& [\%]
\\
\hline

Tungsten
& 1.88
& $1.93\times10^{4}$
& $1.16\times10^{-14}$
& $4.16\times10^{3}$
& $2.82\times10^{-17}$
& 0.244
\\

SUS304
& 0.774
& $7.93\times10^{3}$
& $4.75\times10^{-15}$
& $1.71\times10^{3}$
& $1.16\times10^{-17}$
& 0.244
\\

A5083
& 0.260
& $2.66\times10^{3}$
& $1.59\times10^{-15}$
& $5.73\times10^{2}$
& $3.93\times10^{-18}$
& 0.247
\\

\hline
\end{tabular}

\end{table}

\section{Systematic uncertainty and error propagation}

The GCal provides a calibration torque whose amplitude is determined by
a set of geometric parameters, mass parameters, the moment of inertia of
the torsion bar, and the gravitational constant.
In an ideal configuration, the rotation axis of the GCal rotor is assumed
to coincide with the symmetry center of the torsion bar.
In practice, however, a finite transverse offset between these two axes
is unavoidable and introduces an additional systematic contribution to
the gravitational coupling.
Such an offset modifies the effective separation between the interacting
masses and therefore changes both the magnitude and angular dependence of
the gravitational interaction, leading to a corresponding change in the
predicted amplitude of the calibration line.

Systematic calibration uncertainties are known to directly propagate into
biases in reconstructed detector responses and astrophysical parameter
estimation in gravitational-wave measurements
~\cite{Cahillane2017_Calibration,Abbott2017Calibration}.
In the case of CHRONOS, uncertainties in the physical parameters of the
gravitational calibrator and the torsion-bar system lead to a bias in the
predicted strain-equivalent calibration amplitude,
$h_{\rm GCal}$.
A quantitative evaluation of the resulting systematic uncertainty is
therefore required in order to determine the ultimate calibration accuracy
achievable with the CHRONOS interferometer.

The predicted GCal-induced strain-equivalent response can be written as
\[
h_{\rm GCal}
=
h_{\rm GCal}
(M, m, I, a, b, h, \rho_\perp, G),
\]
where the dependence arises through the gravitational torque and the
mechanical response of the torsion-bar system.

If each parameter contains a small deviation from its nominal value,
\[
M \to M + \delta M, \quad
m \to m + \delta m, \quad
I \to I + \delta I,
\]
\[
a \to a + \delta a, \quad
b \to b + \delta b, \quad
h \to h + \delta h,
\]
\[
\rho_\perp \to \rho_\perp + \delta \rho_\perp, \quad
G \to G + \delta G,
\]
the resulting first-order correction to the calibration amplitude can be
expressed through linear error propagation as
\begin{equation}
\begin{aligned}
\delta h_{\rm GCal}
\simeq {}&
\frac{\partial h_{\rm GCal}}{\partial M}\,\delta M
+\frac{\partial h_{\rm GCal}}{\partial m}\,\delta m
+\frac{\partial h_{\rm GCal}}{\partial I}\,\delta I
+\frac{\partial h_{\rm GCal}}{\partial a}\,\delta a \\
&+\frac{\partial h_{\rm GCal}}{\partial b}\,\delta b
+\frac{\partial h_{\rm GCal}}{\partial h}\,\delta h
+\frac{\partial h_{\rm GCal}}{\partial \rho_\perp}\,\delta \rho_\perp
+\frac{\partial h_{\rm GCal}}{\partial G}\,\delta G .
\end{aligned}
\label{eq:dhgcal_general}
\end{equation}

In the high-frequency regime relevant for CHRONOS,
the strain-equivalent response scales inversely with the moment of inertia,
$h_{\rm GCal} \propto I^{-1}$.
Uncertainty in $I$ therefore introduces a multiplicative systematic
error in the calibration amplitude.
Accurate determination of the torsion-bar moment of inertia is thus
essential for achieving high absolute calibration accuracy,
consistent with the requirements identified for interferometric
gravitational-wave detectors
~\cite{Hall2017CalibrationRequirements}.

The offset term introduces an additional geometric sensitivity that is
absent in the ideally aligned configuration.
In the present formulation, the offset enters through the effective
separation defined in Eq.~\eqref{eq:R_zero} and therefore modifies the
strain-equivalent calibration amplitude through the geometric dependence
of the gravitational interaction.
Because of the rotational symmetry of the system, the leading-order
contribution of the offset appears through $\rho_\perp^2$, implying that
small misalignments primarily affect the calibration amplitude via changes
in the effective distance rather than through a linear directional effect.
Accurate measurement and control of the relative alignment between the
GCal rotation axis and the torsion-bar center are therefore essential for
minimizing systematic errors in the calibration process.

At the calibration frequency, the mechanical susceptibility is fixed by
the detector dynamics and is independent of the GCal parameters.
The derivatives with respect to the parameter set
\[
\Theta = \{M, m, I, a, b, h, \rho_\perp, G\}
\]
therefore reduce to

\begin{equation}
\frac{\partial h_{\rm GCal}}{\partial \Theta_i}
=
\frac{\partial}{\partial \Theta_i}
\left(
\frac{ |\chi_{HF}(2\omega)| \eta_{GCAL} \tau_2}{\eta_g |F_{\rm eff}|\, }
\right),
\qquad
\Theta_i \in \Theta ,
\label{eq:dhgcal_factorization}
\end{equation}
which separates the mechanical response from the parameter dependence of
the gravitational torque amplitude.
This factorization is commonly employed in calibration analyses of
gravitational-wave detectors, where the sensing response and actuator
model are treated independently
~\cite{Cahillane2017_Calibration,Abbott2017Calibration}.

For the moment of inertia, the derivative is obtained as
\begin{equation}
\frac{\partial h_{\rm GCal}}{\partial I}
=
-\frac{h_{\rm GCal}}{I},
\label{eq:I_partial}
\end{equation}
reflecting the purely inertial scaling of the high-frequency response.
The $2f$ torque amplitude is proportional to $GmM$, and therefore
\begin{equation}
\frac{\partial h_{\rm GCal}}{\partial M}
= \frac{h_{\rm GCal}}{M} , \qquad
\frac{\partial h_{\rm GCal}}{\partial m}
= \frac{h_{\rm GCal}}{m}, \qquad
\frac{\partial h_{\rm GCal}}{\partial G}
= \frac{h_{\rm GCal}}{G}.
\label{eq:mass_G_partials}
\end{equation}
These contributions scale directly with the calibration amplitude and
typically lead to subdominant systematic errors due to the high precision
achievable in mass measurements.
In contrast, the offset parameter $\rho_\perp$ modifies the geometrical
dependence of the gravitational interaction through the effective
separation $R_0$ defined above, and therefore affects the
strain-equivalent calibration amplitude through spatial derivatives of
the gravitational torque rather than as a simple multiplicative factor.
Because of the rotational symmetry of the system, the leading-order effect
of the offset appears through $\rho_\perp^2$, implying that small
misalignments primarily modify the calibration amplitude through changes
in the effective separation rather than through a linear directional
effect.
Accurate measurement and control of the relative alignment between the
GCal rotation axis and the torsion-bar center are therefore essential for
minimizing systematic errors in the calibration process.

The strain-equivalent calibration amplitude introduced by the
gravitational calibrator is determined by the gravitational torque
amplitude $\tau_2$ through the relation derived in the previous section.
The $2f$ gravitational torque amplitude can be written compactly as
\[
\tau_2=\frac{G m M}{R_0}\,S(\xi),
\]
where the function $S(\xi)$ describes the higher-order gravitational
coupling arising from the multipole expansion of the interaction and is
written as
\[
S(\xi)=\sum_{n=1}^{N} C_n \,\xi^{2n},
\qquad
S'(\xi)=\sum_{n=1}^{N} 2n\,C_n\,\xi^{2n-1}.
\]
Here the coefficients
\[
C_n =
\left(
\frac{\binom{4n}{\,2n}\binom{2n}{\,n-1}}{2^{6n-4}}
\right)
\]
encode higher-order quadrupole contributions beyond the leading-order
interaction.

Since the strain-equivalent response satisfies
$h_{\rm GCal} \propto \tau_2/I$ in the high-frequency regime,
the offset dependence enters through the geometric derivatives of
$R_0$ and $\xi$ appearing in the multipole expansion above.
Consequently, uncertainties in $\rho_\perp$ modify the calibration
amplitude primarily through changes in the effective separation and the
resulting higher-order gravitational coupling terms.

Using logarithmic differentiation, the parameter dependence of the
gravitational torque amplitude can be written as
\[
\frac{1}{\tau_2}\frac{\partial \tau_2}{\partial \Theta_i}
= -\frac{1}{R_0}\frac{\partial R_0}{\partial \Theta_i}
+ \frac{1}{S(\xi)}\frac{\partial S(\xi)}{\partial \Theta_i},
\]
with
\[
\frac{\partial R_0}{\partial \Theta_i}
= \frac{\Theta_i}{R_0},
\qquad
\Theta_i \in \{a,b,h,\rho_\perp\},
\]
and
\begin{align}
\frac{\partial\xi}{\partial a}
&=\frac{2b}{R_0^2}\left(1-\frac{2a^2}{R_0^2}\right),\\[4pt]
\frac{\partial\xi}{\partial b}
&=\frac{2a}{R_0^2}\left(1-\frac{2b^2}{R_0^2}\right),\\[4pt]
\frac{\partial\xi}{\partial h}
&=-\frac{2h\,\xi}{R_0^2},\\[4pt]
\frac{\partial\xi}{\partial \rho_\perp}
&=-\frac{2\rho_\perp\,\xi}{R_0^2}.
\end{align}

Since the strain-equivalent calibration amplitude satisfies
$h_{\rm GCal} \propto \tau_2/I$ in the high-frequency regime,
the full geometric sensitivities become
\begin{align}
\frac{\partial h_{\rm GCal}}{\partial a}
&= h_{\rm GCal}\left[
   -\frac{a}{R_0^{2}}
   + \frac{S'(\xi)}{S(\xi)}
     \frac{2b}{R_0^{2}}\left(1-\frac{2a^{2}}{R_0^{2}}\right)
   \right], \\[6pt]
\frac{\partial h_{\rm GCal}}{\partial b}
&= h_{\rm GCal}\left[
   -\frac{b}{R_0^{2}}
   + \frac{S'(\xi)}{S(\xi)}
     \frac{2a}{R_0^{2}}\left(1-\frac{2b^{2}}{R_0^{2}}\right)
   \right], \\[6pt]
\frac{\partial h_{\rm GCal}}{\partial h}
&= h_{\rm GCal}\left[
   -\frac{h}{R_0^{2}}
   - \frac{2h\,\xi}{R_0^{2}}\frac{S'(\xi)}{S(\xi)}
   \right], \\[6pt]
\frac{\partial h_{\rm GCal}}{\partial \rho_\perp}
&= h_{\rm GCal}\left[
   -\frac{\rho_\perp}{R_0^{2}}
   - \frac{2\rho_\perp\,\xi}{R_0^{2}}\frac{S'(\xi)}{S(\xi)}
   \right].
\label{eq:geom_partials}
\end{align}

The systematic uncertainty in the predicted calibration amplitude is
estimated by the quadrature sum of the individual contributions,
assuming that the parameter uncertainties are statistically independent:
\begin{equation}
\begin{aligned}
\sigma_{\rm sys}^2
&= \sum_{\Theta_i \in \Theta}
\left(
\frac{\partial h_{\rm GCal}}{\partial \Theta_i}
\,\delta \Theta_i
\right)^2 \\[2pt]
&=
\left(\frac{\partial h_{\rm GCal}}{\partial M}\,\delta M\right)^{2}
+\left(\frac{\partial h_{\rm GCal}}{\partial m}\,\delta m\right)^{2} \\[2pt]
&\quad+
\left(\frac{\partial h_{\rm GCal}}{\partial I}\,\delta I\right)^{2}
+\left(\frac{\partial h_{\rm GCal}}{\partial a}\,\delta a\right)^{2} \\[2pt]
&\quad+
\left(\frac{\partial h_{\rm GCal}}{\partial b}\,\delta b\right)^{2}
+\left(\frac{\partial h_{\rm GCal}}{\partial h}\,\delta h\right)^{2} \\[2pt]
&\quad+
\left(\frac{\partial h_{\rm GCal}}{\partial \rho_\perp}\,\delta \rho_\perp\right)^{2}
+\left(\frac{\partial h_{\rm GCal}}{\partial G}\,\delta G\right)^{2}.
\end{aligned}
\label{eq:sys_total}
\end{equation}

This formulation follows the standard linear error-propagation framework
commonly adopted in calibration analyses of gravitational-wave detectors
~\cite{Cahillane2017_Calibration,Abbott2017Calibration}.

The uncertainty in the quadrupole mass distribution is limited by the
precision of the electronic balance used for mass measurements.
In this study, the rotor masses are assumed to be made of tungsten,
SUS304 stainless steel, and A5083 aluminum alloy.
The masses is measured by using a GC-6000 electronic balance, and its
tolerance is assumed to be $0.2\,\mathrm{g}$~\cite{CG6000}.

The rotor disk can be fabricated using numerically controlled (NC)
milling~\cite{Inoue2016_ARCoating}.
This machining method typically achieves dimensional accuracies better
than $20\,\mu\mathrm{m}$.
The torsion-bar test masses are planned to be assembled using
hydroxy-catalysis bonding, for which a positioning accuracy of
approximately $50\,\mu\mathrm{m}$ is assumed.
Geometrical measurements are assumed to be performed using a coordinate
measuring machine (CMM), whose accuracy is taken to be
$2\,\mu\mathrm{m}$~\cite{Inoue2016_ARCoating}.
This indicates that both the rotor geometry and mass distribution can be
characterized with sufficiently small uncertainty using CMM measurements.
The geometry of the test masses can likewise be measured using the same
method.
The relative distance between the rotor and the test mass, as well as the
offset of their central axes, can in principle be measured with an
accuracy of about $100\,\mu\mathrm{m}$ using a three-dimensional
measuring system; however, a conservative margin is adopted in this work,
and an uncertainty of $1\,\mathrm{mm}$ is assumed.
For the gravitational constant, the current experimental uncertainty is
adopted.

We use the CODATA 2022 recommended value
of the gravitational constant
$G = 6.67430(15)\times10^{-11}\,\mathrm{m^3\,kg^{-1}\,s^{-2}}$~\cite{CODATA2022}.

The systematic uncertainty was evaluated using a first-order perturbation
framework based on analytical parameter derivatives.
The resulting fractional systematic error,
\[
\frac{\sigma_{\rm sys}}{h_{\rm GCal}}
\simeq 2.4\times10^{-3}
\;(0.24\%),
\]
is dominated by geometric uncertainties in the rotor--sensor alignment,
whereas uncertainties associated with the calibration masses and the
gravitational constant remain subdominant.

In addition to these experimental uncertainties,
the use of the high-frequency approximation
\(
\chi(\Omega)\simeq -1/(I\Omega^2)
\)
introduces an additional frequency-dependent systematic deviation.
As discussed in Appendix~C, the deviation between the exact transfer
function and the high-frequency approximation for the SUS-steel suspension
is approximately given by
\[
\left|
\frac{\chi_{\rm exact}}{\chi_{\rm HF}}-1
\right|
\simeq
0.16\%
\left(
\frac{0.1\,{\rm Hz}}{f}
\right)^2 .
\]

Therefore, this correction should be included when the high-frequency
approximation is adopted in the calibration analysis.
For most calibration-line injections performed well above
\(0.1\,{\rm Hz}\),
the induced deviation is sufficiently smaller than the nominal
systematic uncertainty and can be safely neglected.
However, near the lower edge of the observational band around
\(0.1\,{\rm Hz}\),
the deviation becomes comparable to the quoted calibration uncertainty.
In such cases, it is preferable to evaluate the detector response
using the full exact transfer function rather than the asymptotic
high-frequency approximation.

\section{Discussion}
\label{sec:discussion}

\subsection{Force-coupled and torque-coupled GCals}
\label{sec:length_vs_torque}

The essential difference between force-coupled and torque-coupled GCals
lies in how the gravitational modulation couples to the torsional degree
of freedom.
In conventional force-coupled configurations,
the gravitational forces acting on both ends of the torsion bar are
largely aligned, leading to partial cancellation of the induced torque.
As a result, the rotational response becomes weak, particularly in the
low-frequency regime.

In contrast, the torque-coupled GCal proposed in this work places the
rotating masses directly beneath the torsion bar so that a gravitational
torque is directly applied to the test mass.
This configuration avoids cancellation of moments and enables efficient
coupling to the rotational degree of freedom, allowing higher SNR to be
maintained at low frequencies under otherwise identical conditions.
Furthermore, the proposed scheme can operate with a single GCal while
maintaining sub-percent-level systematic uncertainty.

These results demonstrate that torque-coupled GCal provides a viable
pathway toward absolute calibration in the sub-Hz frequency band,
where conventional radiation-pressure-based calibration methods become
less effective due to reduced mechanical response and increased
technical noise.
Because the calibration signal is generated gravitationally,
it naturally couples to the most sensitive degree of freedom and remains
well defined independently of optical gain fluctuations or interferometer
control dynamics.

The combination of high SNR, analytically tractable systematics,
and reduced dependence on auxiliary control models makes the
torque-coupled GCal particularly suitable for next-generation
low-frequency detectors such as CHRONOS, TOBA, and TorPeDO.
Overall, the method presented here provides a clean,
analytically tractable, and experimentally feasible pathway toward
absolute calibration of torsion-bar interferometers and future
sub-Hz GW detectors.

\subsection{Validation of the quadrupole approximation}

To validate the analytical quadrupole description used throughout this
work, we compared the analytical expression with an exact numerical
evaluation of the full four-body Newtonian interaction.
The exact torque was computed directly from the Newtonian potential
without using the multipole expansion, and the
\(2f_{\rm rot}\)
component was extracted numerically by Fourier decomposition.

For the nominal geometry, the relative difference between the analytical
quadrupole model and the exact numerical result was found to be
approximately
\(10^{-6}\),
corresponding to a discrepancy of only
\(10^{-4}\%\).
This demonstrates that the truncation of the analytical expansion does
not introduce a significant error for the geometry considered in this
work.

The remaining modeling uncertainty is therefore expected to be dominated
not by the multipole truncation itself, but by practical uncertainties
associated with the detailed detector geometry and calibration
parameters.

\subsection{Experimental validation of non-gravitational coupling}

In practical implementations of gravitational calibrators,
non-gravitational coupling mechanisms must also be carefully evaluated.
Possible contamination channels include magnetic coupling,
vibrational injection, scattered light, electrical pickup,
and acoustic coupling associated with the rotating actuator system.
Although such effects are expected to be small in the present geometry,
experimental validation is essential for establishing the robustness of
the calibration.

Similar validation strategies have been developed in existing
gravitational calibrator systems.
For example, the Virgo Newtonian calibrator (NCal) performed dedicated
zero-injection measurements by rotating the actuator in configurations
designed to suppress the gravitational signal while preserving possible
environmental couplings.

In the context of the present torque-coupled GCal scheme,
similar validation procedures can be implemented.
One possible approach is a vertical-rotation configuration in which the
rotor continues to generate environmental disturbances associated with
the motor system while strongly suppressing the Newtonian torque acting
on the torsion bar.
Such measurements would provide a direct estimate of residual
non-gravitational coupling.

\subsection{Future validation strategy}

Additional null tests can also be performed using dummy rotors with
identical mechanical properties but reduced mass density, allowing the
separation of gravitational and environmental contributions.
Furthermore, comparison between symmetric and intentionally asymmetric
rotor configurations may provide an additional consistency check of the
quadrupole-induced calibration signal.

Future experimental studies will therefore focus on the quantitative
validation of possible non-gravitational couplings and the development
of dedicated null-injection measurements.
Such tests will be important for establishing the reliability of
sub-percent-level calibration using torque-coupled gravitational field
calibrators.

Possible future validation procedures include:
\begin{itemize}
\item vertical rotation tests,
\item symmetric null configurations,
\item dummy rotor measurements,
\item comparison between multiple rotor geometries,
\item environmental coupling measurements with suppressed
gravitational signals.
\end{itemize}

These studies will help establish an experimentally validated systematic
uncertainty budget for future high-precision gravitational calibration
systems.

\section{Conclusion}
\label{sec:conclusion}

We have calculated the first demonstration and systematic formulation of a
torque-coupled GCal designed for sub-Hz
precision measurements with a torsion-bar test mass.
Unlike conventional force-coupled GCals used in kilometer-scale
interferometers, the present approach couples the Newtonian gravitational
field directly to the rotational degree of freedom of the detector.
This configuration provides a calibration scheme that is intrinsically
less sensitive to suspension dynamics and interferometer control loops,
thereby enabling a conceptually simpler and more direct calibration
pathway for torsion-based gravitational-wave detectors.

An analytical framework describing the gravitational interaction between
a rotating quadrupole rotor and a torsion bar was developed using a
systematic multipole expansion.
The resulting closed-form expression for the torque amplitude allows
direct evaluation of higher-order gravitational contributions as well as
analytic computation of parameter derivatives.
This enables transparent propagation of systematic uncertainties and
provides a reproducible calibration procedure without relying on numerical
potential-field integration.

Using parameters corresponding to the current CHRONOS configuration, we
demonstrated that the GCal-induced calibration signal appears as a narrow
spectral line at $2f_{\rm rot}$ within the operational band of
$0.1$--$10~\mathrm{Hz}$.
Because the torsion-bar resonance frequency
($\omega_0/2\pi < 0.01~\mathrm{Hz}$) lies well below the calibration
band, the detector response is dominated by inertial motion, allowing a
direct conversion between the applied gravitational torque and the
strain-equivalent detector response.

A central result of this study is the explicit dependence of the
calibration performance on the mass density of the rotor material.
Since the gravitational torque scales linearly with the calibration mass,
the achievable strain-equivalent calibration amplitude and the
corresponding signal-to-noise ratio density,
defined as $|h_{\rm GCal}|/S_h$,
are primarily determined by the material density when the rotor geometry
is fixed.
For identical geometrical configurations, the calibration-line SNR density
at $1\,\mathrm{Hz}$ reaches
$\mathrm{SNR}_{\rm tungsten}=4.16\times10^{3}$,
$\mathrm{SNR}_{\rm SUS304}=1.71\times10^{3}$, and
$\mathrm{SNR}_{\rm A5083}=5.73\times10^{2}$.
The corresponding strain-equivalent calibration amplitudes are
$1.16\times10^{-14}$,
$4.75\times10^{-15}$, and
$1.59\times10^{-15}$, respectively.
This clear density scaling demonstrates that high-density materials
substantially enhance calibration margin while preserving the same
mechanical configuration, providing a practical design guideline for
future implementations.

The systematic uncertainty was evaluated using a first-order perturbation
framework based on analytical parameter derivatives.
The resulting fractional systematic error,
\[
\frac{\sigma_{\rm sys}}{h_{\rm GCal}}
\simeq 2.4\times10^{-3}
\;(0.24\%),
\]
is dominated by geometric uncertainties in the rotor--sensor alignment,
whereas uncertainties associated with the calibration masses and the
gravitational constant remain subdominant.
In absolute terms, the corresponding systematic uncertainty of the
strain-equivalent calibration amplitude at $1\,\mathrm{Hz}$ is of order
$10^{-17}$, depending on the rotor material.
This level of accuracy establishes the achievable absolute calibration
limit of the torque-coupled GCal under realistic experimental conditions.

Overall, the results presented here establish torque-coupled gravitational
calibration as a practical and scalable method for torsion-bar
gravitational-wave detectors.
The combination of a large calibration-line amplitude relative to the
detector noise, analytically tractable systematics, and reduced dependence
on interferometer control models provides a robust calibration strategy
for CHRONOS and future sub-Hz gravitational-wave observatories.
More broadly, the method demonstrates that gravity-based calibration can
serve as a key metrological tool for next-generation low-frequency
detectors targeting precision astrophysics and cosmology.

\section*{Acknowledgment}

We thank Masashi Hazumi for valuable academic advice during the preparation of this manuscript.
We also thank Masaya Hasegawa, Takahiro Kanayama, Satoki Matsushita, Hirokazu Murakami, Mai Inoue, Ryuji Shibuya, and Chia-Ming Kuo for their support in establishing the CHRONOS team.

Y.~Inoue acknowledges support from the National Science and Technology Council (NSTC), the Center for High Energy and High Field Physics (CHiP), and Academia Sinica in Taiwan under Grant Nos.~114-2112-M-008-006 and AS-TP-112-M01.


\appendix
\section*{Appendix}
\section*{A. Ordering Scheme and Offset-Suppressed Corrections}
\label{app:ordering}

In this appendix, we summarize the perturbative ordering scheme
used in the analytical derivation of the gravitational calibration
torque and quantify the effect of the neglected offset-dependent
terms.

The geometry is characterized by the hierarchy $\rho \ll a,b \ll R_0$,
where
$\rho$
is the lateral offset,
$a$
and
$b$
are the characteristic arm lengths of the torsion bar and rotating
source masses, respectively, and
$R_0$
is the mean source--detector separation.

According to Eq.~\eqref{eq:approx},
\begin{equation}
\frac{\alpha}{\xi}
=
\mathcal{O}\!\left(\frac{\rho}{b}\right),
\qquad
\frac{\beta}{\xi}
=
\mathcal{O}\!\left(\frac{\rho}{a}\right),
\end{equation}
so that
\begin{equation}
\alpha,\beta \ll \xi
\qquad
(\rho \ll a,b).
\end{equation}

The analytical derivation retains the leading symmetric contributions
proportional to
$\xi^{2n}$.
Terms involving
$\alpha$
and
$\beta$
beyond linear order are neglected.
The linear contributions cancel identically in the symmetric
four-mass configuration, while the remaining offset-dependent terms are
suppressed relative to the leading contribution by
\begin{equation}
\mathcal{O}\!\left(\frac{\rho}{\min(a,b)}\right).
\end{equation}

For the nominal parameters

$a=0.302~{\rm m},
b=0.160~{\rm m},
h=2.0~{\rm m},
\rho=1~{\rm mm},$
we obtain
$R_0
=
\sqrt{a^2+b^2+h^2+\rho^2}
\simeq
2.029~{\rm m},$
and therefore
$\xi
\simeq
2.35\times10^{-2}.$

Taking the conservative bounds
\(
|C_\phi|,|C_\theta|\leq \rho
\),
the offset parameters satisfy
\begin{equation}
\alpha_{\rm max}
\simeq
1.47\times10^{-4},
\qquad
\beta_{\rm max}
\simeq
7.77\times10^{-5},
\end{equation}
which gives
\begin{equation}
\frac{\alpha_{\rm max}}{\xi}
\simeq
6.25\times10^{-3},
\qquad
\frac{\beta_{\rm max}}{\xi}
\simeq
3.31\times10^{-3}.
\end{equation}

To verify that the neglected offset-dependent terms do not affect
the calibration line, we numerically evaluated the full Newtonian
potential without setting
$\alpha=\beta=0$
and extracted the
$2f_{\rm rot}$
Fourier component of the torque.
We define
\begin{equation}
\delta_\rho
\equiv
\frac{
\tau_{2f}^{\rm full}(\rho)
-
\tau_{2f}^{(0)}
}{
\tau_{2f}^{(0)}
},
\end{equation}
where
$\tau_{2f}^{(0)}$
denotes the symmetric
$(\rho=0)$
result.

For the nominal parameters, we obtain
\begin{equation}
\tau_{2f}^{(0)}
=
4.382057\times10^{-12}\ {\rm N\,m},
\end{equation}
and
\begin{equation}
\tau_{2f}^{\rm full}
=
4.382049\times10^{-12}\ {\rm N\,m}.
\end{equation}

Thus the offset-induced correction to the
$2f_{\rm rot}$
calibration amplitude is only
\begin{equation}
|\delta_\rho|
=
1.7\times10^{-4}\%,
\end{equation}
which is substantially smaller than the other systematic uncertainties
considered in this work.


\section*{B. Finite-width correction to the GCal quadrupole response}
\label{app:eta_gcal}
{ \color{black}
In the ideal thin-bar approximation, the leading quadrupole response of
the torsion bar is characterized solely by the effective longitudinal
lever arm $a$.

For a realistic torsion bar with finite transverse extent, the mass
distribution generates an additional quadrupole contribution associated
with the effective width $w$. The longitudinal and transverse
contributions are denoted by $\tau_{2,a}$ and $\tau_{2,w}$,
respectively. The latter enters with the opposite sign because the
transverse mass distribution reduces the net quadrupole moment.

The effective calibration torque is therefore written as
\begin{equation}
\tau_{2}^{\rm eff}
=
\tau_{2,a}
-
\tau_{2,w},
\label{eq:tau_eff_appendix}
\end{equation}
where $\tau_{2,a}$ represents the quadrupole torque arising from the
effective arm length $a$, while $\tau_{2,w}$ represents the finite-width
correction associated with the transverse extent of the torsion-bar mass
distribution.

Because the torsion-bar mass distribution is symmetric under the
reflections \(x\rightarrow -x\) and \(y\rightarrow -y\), the mixed
quadrupole component vanishes,
\begin{equation}
Q_{xy}
=
\int \rho\,xy\,dV
=
0.
\end{equation}
Accordingly, the leading correction appears through the difference of
the diagonal quadrupole components,
\begin{equation}
Q_{xx}-Q_{yy}
\propto
a^2-w^2.
\end{equation}

Using the multipole expansion derived in Eq.~\eqref{eq:tau2_expand}, we define the geometrical correction factor
\begin{equation}
\eta_{\rm GCal}
\equiv
\frac{
\tau_{2,a}-\tau_{2,w}
}{
\tau_{2,a}
}.
\label{eq:eta_gcal_def_appendix}
\end{equation}

At leading order,
\begin{equation}
\eta_{\rm GCal}
\simeq
1-\frac{w^2}{a^2}.
\label{eq:eta_leading_appendix}
\end{equation}

For the present geometry,
\begin{equation}
a=0.302\,{\rm m},
\qquad
w=0.050\,{\rm m},
\qquad
R_0=2.029\,{\rm m},
\end{equation}

We define the dimensionless parameters

\begin{equation}
\xi_a=\frac{2ab}{R_0^2},
\qquad
\xi_w=\frac{2wb}{R_0^2},
\end{equation}

which are consistent with the definition
$\xi = 2ab/R_0^2$ used in the main text.

For the present geometry,

\begin{equation}
\xi_a =0.0235,
\qquad
\xi_w =0.00389.
\end{equation}

Using Eq.~(\ref{eq:tau2_expand}) up to
\(\mathcal O(\xi^6)\), we obtain
\begin{align}
\tau_{2,a}
&=
7.530\times10^{-10}\ {\rm N\,m},
\\
\tau_{2,w}
&=
1.934\times10^{-11}\ {\rm N\,m},
\\
\tau_2^{\rm eff}
&=
7.337\times10^{-10}\ {\rm N\,m}.
\end{align}

The resulting correction factor becomes
\begin{equation}
\eta_{\rm GCal}
=
0.9743.
\end{equation}

For comparison, the leading-order approximation gives
\begin{equation}
1-\frac{w^2}{a^2}
=
0.9726.
\end{equation}
The difference between the leading-order estimate and the
\(\mathcal O(\xi^6)\) result is therefore only
\begin{equation}
\frac{
\eta_{\rm GCal}^{(\xi^6)}
-
\eta_{\rm GCal}^{(\xi^2)}
}{
\eta_{\rm GCal}^{(\xi^2)}
}
\simeq
1.8\times10^{-3},
\end{equation}
corresponding to a \(0.18\%\) correction.

This demonstrates that the dominant finite-width effect is already
captured by the leading quadrupole scaling
\((1-w^2/a^2)\), while higher-order multipole corrections remain
subdominant for the present geometry.


\section*{C. Validity of the High-Frequency Approximation}
\label{app:transfer_function}

In the main text, the mechanical response of the torsion-bar detector
was approximated by Eq.~\eqref{eq:approx_HF}, which is valid in the regime
\(
\Omega \gg \Omega_0
\),
where
\(
\Omega_0
\)
is the torsional resonance frequency.

The exact torsional response function is obtained as Eq.~\eqref{eq:chi_general}.

The torsional resonance frequency is determined by the torsional spring
constant
\(\mu\)
through
\begin{equation}
\Omega_0
=
\sqrt{\frac{\mu}{I}},
\qquad
f_0
=
\frac{1}{2\pi}
\sqrt{\frac{\mu}{I}},
\label{eq:torsion_resonance}
\end{equation}
where
\(I\)
is the moment of inertia of the torsion bar.

For a cylindrical suspension wire of radius
\(r\)
and length
\(l\),
the torsional spring constant is
\begin{equation}
\mu
=
\frac{\pi G r^4}{2l},
\label{eq:torsion_constant}
\end{equation}
where
\(G\)
is the shear modulus.
The shear modulus is related to the Young modulus
\(E\)
and Poisson ratio
\(\nu\)
through
\begin{equation}
G
=
\frac{E}{2(1+\nu)}.
\label{eq:shear_modulus}
\end{equation}

Table~\ref{tab:suspension_materials}
summarizes representative material parameters used in the comparison.

\begin{table}[t]
\centering
\caption{
Representative suspension parameters used for the transfer-function
comparison shown in
Fig.~\ref{fig:transfer_function_deviation}.
The calculations assume
\(I=19.9~{\rm kg\,m^2}\),
wire radius
\(r=1~{\rm mm}\),
and suspension length
\(l=10~{\rm m}\).
}
\small
\begin{tabular}{cccc}
\hline\hline
Material
& $Q$
& $G$ [GPa]
& $f_0$ [Hz]
\\
\hline

SUS (Steel)
& $10^5$
& 77.5
& $3.94\times10^{-3}$ \\

Silicon
& $10^6$
& 50.8
& $3.19\times10^{-3}$ \\

Sapphire
& $10^7$
& 163
& $5.70\times10^{-3}$ \\

\hline\hline
\end{tabular}
\label{tab:suspension_materials}
\end{table}

\begin{figure}[t]
\centering
\includegraphics[width=0.95\linewidth]
{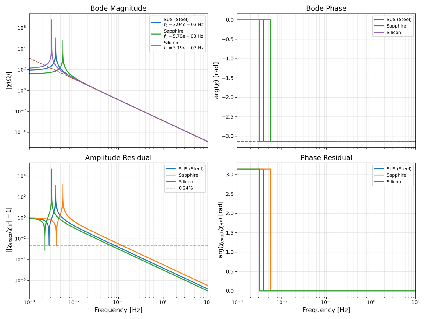}
\caption{
Bode comparison between the exact complex torsional transfer function
and the high-frequency approximation
$\chi_{\rm HF}(\Omega)\simeq -1/(I\Omega^2)$
for representative suspension materials.
The upper panels show the transfer-function magnitude and phase,
while the lower panels show the corresponding amplitude and phase residuals.
The dashed horizontal line in the amplitude residual panel
indicates the nominal systematic uncertainty level of
\(0.24\%\)
adopted in this work.
Although the resonance frequencies differ depending on the suspension
material, all transfer functions asymptotically converge to the same
complex inertial response proportional to
\(1/\Omega^2\)
with a phase shift of
\(-\pi\).
}
\label{fig:transfer_function_deviation}
\end{figure}

We evaluate the deviation from the high-frequency approximation
using the complex fractional residual defined as
\begin{equation}
\frac{\delta\chi(\Omega)}{\chi_{\rm HF}(\Omega)}
=
\frac{
\chi_{\rm exact}(\Omega)
}{
\chi_{\rm HF}(\Omega)
}
-1 .
\end{equation}

The magnitude and phase residuals shown in
Fig.~\ref{fig:transfer_function_deviation}
are respectively given by
\begin{equation}
\left|
\frac{\delta\chi(\Omega)}{\chi_{\rm HF}(\Omega)}
\right|,
\end{equation}
and
\begin{equation}
\arg
\left(
\frac{
\chi_{\rm exact}(\Omega)
}{
\chi_{\rm HF}(\Omega)
}
\right).
\end{equation}

Figure~\ref{fig:transfer_function_deviation}
shows that both the magnitude and phase of the exact transfer function
converge toward the high-frequency approximation above the resonance
frequency.

For the SUS-steel suspension, the asymptotic deviation
between the exact transfer function and the high-frequency
approximation is well described by
\begin{equation}
\left|
\frac{\chi_{\rm exact}}{\chi_{\rm HF}}-1
\right|
\simeq
0.16 \%
\left(
\frac{0.1\,{\rm Hz}}{f}
\right)^2 .
\end{equation}

Using the nominal systematic uncertainty adopted in this work,
the crossover frequency is obtained from
\begin{equation}
f_{\rm cross}
\simeq
8.0\times10^{-2}\ {\rm Hz}.
\end{equation}

Therefore, above approximately
\(0.08\,{\rm Hz}\),
the error introduced by the high-frequency approximation
becomes smaller than the nominal calibration systematic uncertainty.
This indicates that the approximation
\(
\chi(\Omega)\simeq -1/(I\Omega^2)
\)
is sufficiently accurate in the observational frequency band
relevant to the present study.
}

\acknowledgments
We thank Masashi Hazumi for valuable academic advice during the preparation of this manuscript.
We also thank Masaya Hasegawa, Takahiro Kanayama, Satoki Matsushita, Hirokazu Murakami, Mai Inoue, Ryuji Shibuya, and Chia-Ming Kuo for their support in establishing the CHRONOS team.

Y.~Inoue acknowledges support from the National Science and Technology Council (NSTC), the Center for High Energy and High Field Physics (CHiP), and Academia Sinica in Taiwan under Grant Nos.~114-2112-M-008-006 and AS-TP-112-M01.

\bibliographystyle{apsrev4-1}
\bibliography{chronos_GCal}
\end{document}